\documentclass[twocolumn,secnumarabic,amssymb, nobibnotes, aps, prc,nofootinbib,fleqn, floatfix,showpacs,superscriptaddress]{revtex4-1}
\usepackage[utf8]{inputenc}
\usepackage{graphicx}
\usepackage{subfigure}
\usepackage{amsmath,amssymb}
\usepackage{bm}
\usepackage{multirow}

\begin{document}

\title{Empirical residual neutron-proton interaction in odd-odd nuclei}%

\author{Zheying Wu}%
\affiliation{Department of Physics, Royal Institute of Technology (KTH), SE-10691 Stockholm, Sweden}
\author{S. A. Changizi}%
\affiliation{Department of Physics, Royal Institute of Technology (KTH), SE-10691 Stockholm, Sweden}

\author{Chong Qi}%
\email{chongq@kth.se}
\affiliation{Department of Physics, Royal Institute of Technology (KTH), SE-10691 Stockholm, Sweden}

\date{\today}%

\begin{abstract}
Two types of average neutron-proton interaction formulas are compared: In the first type, neutron-proton interactions for even-even and odd-$A$ nuclei extracted from experimental binding energies show a smooth behavior as a function of mass number $A$ and are dominated by the contribution from the symmetry energy. Whereas in the second type large systematic staggering is seen between even-$A$ and odd-$A$ nuclei. This deviation is understood in term of the additional neutron-proton interaction in odd-odd nuclei relative to the neighboring even-even and odd-$A$ systems. We explore three possible ways to extract this additional interaction from the binding energy difference of neighboring nuclei. The  extracted interactions are positive in nearly all cases and show weak dependence on the mass number. The empirical interactions are also compared with theoretical values extracted from recent nuclear mass models where large unexpected fluctuations are seen in certain nuclei. The reproduction of the residual neutron-proton interaction and the correction of those irregular fluctuations can be a good criterion for the refinement of those mass models.

\pacs{21.10.Dr, 21.30.Fe, 21.60.Jz, 24.10.Cn	}
\end{abstract}

\maketitle
\section{Introduction}
\begin{figure}[htdp]
\includegraphics[width=0.4\textwidth]{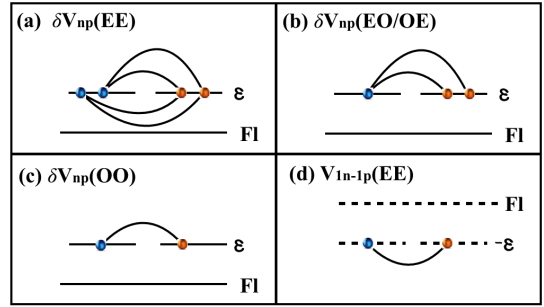}
\caption{\label{shiyi} (Color online) Illustration of the np interaction in even-even (EE) (a), odd-$A$ (EO/OE) (b) and odd-odd (OO) (c) nuclei as extracted from Eqs. (\ref{vpn-ee}-\ref{vpn-oo}) as well as those from Eq. (\ref{vpn-zhao}) for even-even nuclei, which can be seen as a hole-hole-like np interaction.}
\end{figure}

Systematic studies on nuclear masses, in particular their global smooth behaviors and local fluctuations, have revealed rich information about nuclear structure and the underlying effective interaction. It is possible to isolate specific interaction channels  by using appropriate relative mass differences. For example, the pairing interaction can be extracted from the odd-even staggering of nuclear masses (see, e.g., Refs. \cite{Hov13, Sar15,Sar15a, Satu98,Ber09} and references therein). Another prominent example is the so-called average neutron-proton (np) interaction in the even-even nucleus, which can be extracted from the double difference of binding
energies as~\cite{Zhang89},
\begin{eqnarray}\label{vpn-ee}
\nonumber  \delta V_{np} {(Z, N)}&=& \frac{1}{4}\left[
B(Z,N)+B(Z-2,N-2)\right. \\
&&- \left.B(Z-2,N)-B(Z,N-2) \right]
\end{eqnarray}
where $B(Z,N)$ is the (positive) binding energy of a nucleus with $Z$ protons and $N$ neutrons.
The factor $1/4$ takes into account the fact that, as illustrated in the upper-left panel of Fig. \ref{shiyi}, four additional np pair interactions are
formed by the last two protons and neutrons. This filter has been extensively studied in the literature \cite{Cas05,Cas06,Cas06a,Cas06b,Isa95,Satu97,Chen09,
Bon13,Cas09,Chas07,Isa07}. The empirical values from experimental data were compared with those predicted by density functional as well as mass formula calculations \cite{Sto07,Qi12}.

The average np interaction in odd-$A$ and odd-odd
nuclei can be extracted in a similar way as
\begin{eqnarray}\label{vpn-eo}
 && \delta V_{np} {(Z, N-1)}= \frac{1}{2}[
B(Z,N-1)+B(Z-2,N-2) \hspace{5mm}\nonumber\\
&&\hspace{2.2cm}-B(Z-2,N-1)-B(Z,N-2)],
\end{eqnarray}
\begin{eqnarray}\label{vpn-oe}
\nonumber &&  \delta V_{np} {(Z-1, N)}= \frac{1}{2}[
B(Z-1,N)+B(Z-2,N-2)\hspace{5mm} \\
&&\hspace{2.2cm}-B(Z-2,N)-B(Z-1,N-2)],
\end{eqnarray}
which involve two np pairs (Fig. 1b) and
\begin{eqnarray}\label{vpn-oo}
&& \delta V_{np} {(Z-1, N-1)}=\hspace{5cm}\nonumber\\
&&\hspace{1.8cm}\nonumber B(Z-1,N-1)+B(Z-2,N-2)\\
&&\hspace{1.7cm}-B(Z-1,N-2)-B(Z-2,N-1),
\end{eqnarray}
involving one np pair (Fig. 1c). We have assumed that $N$ and $Z$ only take even numbers in Eqs. (1-4).
Empirical studies of nuclear masses suggest that average np interactions for even-even and neighboring odd-$A$ nuclei extracted from experimental data are roughly the same and show  rather smooth behavior as a function of $A$ \cite{Sto07,Qi12}. Whereas values for odd-odd nuclei are systematically larger than those for neighboring even-even and odd-$A$ nuclei. 

In addition to the family of $\delta V_{np}$ relations shown above in Eqs. (1-4), there is another way to extract the average np interaction as 
\begin{eqnarray}\label{vpn-zhao}
  V_{1n-1p} {(\mathcal{Z}, \mathcal{N})}= B(\mathcal{Z},\mathcal{N})+B(\mathcal{Z}-1,\mathcal{N}-1)  \nonumber \\
- B(\mathcal{Z},\mathcal{N}-1)-B(\mathcal{Z}-1,\mathcal{N}), 
\end{eqnarray}
where $\mathcal{Z}$ and $\mathcal{N}$ denote proton and neutron umbers which can take both even and odd values. This was proposed in Ref. \cite{Bas71}
and applied recently in Refs. \cite{zhao15,zhao11,zhao12,zhao13,Zhao14}. 

In this paper we are interested to explore the difference between above two families of np formulas. We will show that the difference can be useful for our understanding of the general properties of the np interaction as well as the additional binding between the last unpaired proton and neutron in odd-odd nuclei. This additional binding can be extracted from experimental data in simple ways. We will also compare those empirical np interactions with theoretical values extracted from recent nuclear mass models. 

\section{Comparison between the two types of np interactions}
\begin{figure}
\includegraphics[width=0.45\textwidth]{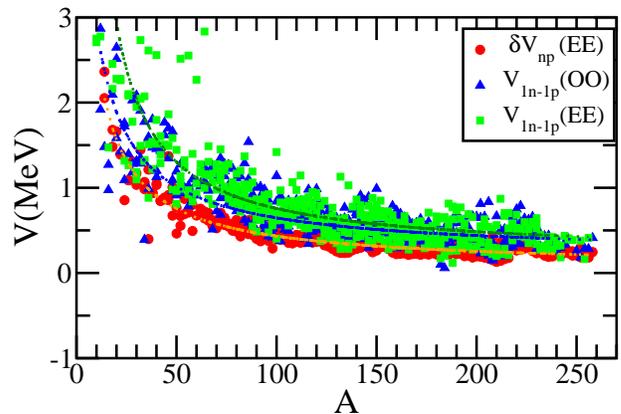}
\caption{\label{ee-oo} (Color online) The average np interactions for even-even and odd-odd nuclei as extracted from experimental data \cite{AM12} by using $V_{1n-1p}$, Eq. (\ref{vpn-zhao}), in comparison to $\delta V_{np}(EE)$ from Eq. (\ref{vpn-ee}). The dotted lines are determined by fitting to the corresponding data points.}
\end{figure}

The average np interaction formulas Eqs. (\ref{vpn-oo}) and (\ref{vpn-zhao}) are the same for odd-odd nuclei. They measure the energy gain by the odd proton and odd neutron relative to the even-even core due to the additional np interaction between the two particles. On the other hand, the np interaction for even-even nuclei from  Eq. (\ref{vpn-zhao}) involves the breaking of two identical pairs, which one intentionally avoided in the construction of the first family of average np interaction. On the first glance, one may suspect that the np interaction for even-even and odd-$A$ nuclei extracted from Eq. (\ref{vpn-zhao}) may be perturbed by the neutron-neutron and proton-proton pairing interactions.
However, it is noticed that these pair energies will cancel each other and will not affect the final np interaction energy if the pairing interactions in even-even nuclei are the same as those of the corresponding odd-$A$ nuclei with one less neutron or proton. This can also be understood from Eq. (\ref{vpn-zhao}): Similar to $V_{1n-1p}(OO)$ which measures the energy gain relative to the even-even core with one less np pair, one can re-interpret $V_{1n-1p}(Z,N)$  for a even-even nucleus as a measure of the np interaction between the $(Z-1)$th proton and $(N-1)$th neutron relative to the even-even nucleus $(Z,N)$ with one more np pair. This is illustrated in the lower-right panel of Fig. \ref{shiyi}. Indeed, as shown in Fig. \ref{ee-oo}, $V_{1n-1p}(EE)$ are roughly the same as those for the odd-odd nuclei with one less np pair.

\begin{figure}
\includegraphics[width=0.45\textwidth]{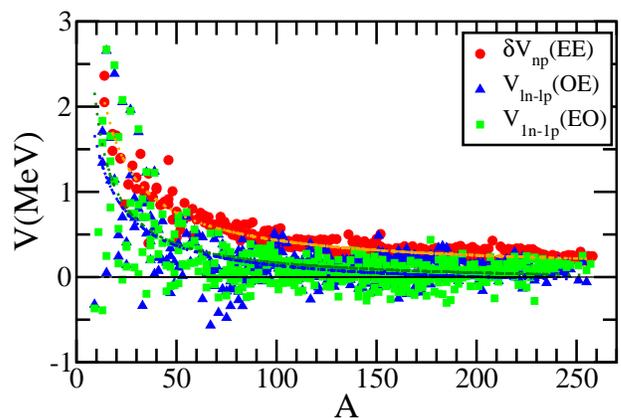}
\caption{\label{eo-oe} (Color online) The average np interactions for odd-$A$ nuclei as extracted from experimental data \cite{AM12} by using $V_{1n-1p}$ in comparison to $\delta V_{np}(EE)$ from Eq. (\ref{vpn-ee}). The dotted lines are determined by fitting to the corresponding data points.}
\end{figure}

Moreover, a striking feature one notice is that the average np interactions $V_{1n-1p}(EE)$ and $V_{1n-1p}(OO)$ shown in Fig. \ref{ee-oo} are systematically larger than $\delta V_{np}(EE)$ for neighboring nuclei. 
In Fig. \ref{eo-oe} we also extracted $V_{1n-1p}$ for even-odd and odd-even nuclei, which in most cases, however, are smaller than $\delta V_{np}(EE)$.
These different deviation behaviors are related to the fact that $\delta V_{np}$ for odd-$A$ nuclei are actually the average of $V_{1n-1p}(EO/OE)$ and $V_{1n-1p}(OO)$ as
\begin{eqnarray}\label{vpn-GKs}
&&\delta V_{np}(Z-1,N)\hspace{6cm}\nonumber\\
&&\hspace{3mm}=\frac{1}{2}[ V_{1n-1p} {(Z-1, N)}+V_{1n-1p} {(Z-1, N-1)}].
\end{eqnarray}
The reason for the systematic deviations seen in Figs. \ref{ee-oo} and \ref{eo-oe} will be analyzed below.

\section{The residual np interaction
in odd-odd nuclei}

From a phenomenological point of view, it is understood that $\delta V_{np}$ for even-even and odd-$A$ nuclei are dominated by contributions from nuclear symmetry energy which is induced by the monopole np interaction (the mean field). We have evaluated the contribution from the standard liquid drop model (with parameters from Ref. \cite{Wu15}) where, as expected, the values for $\delta V_{np}$ and $V_{1n-1p}$ are nearly identical and show a very smooth behavior as a function of $A$.
On the other hand, empirical
 $\delta V_{np}(OO)$ extracted from experimental binding energies can be a mixture of the symmetry energy effect and the re-coupling effect due to the residual np interaction between the two unpaired particles. The mean field effect has to be properly filtered out if one aims at studying the residual np coupling. This is important for our eventual clarification of the role played by np pairing correlation in nuclei (see, e.g., Refs. \cite{Macc00,Qi11,Good01,Satu01}) and
for our understanding of the density functional theory \cite{Ber07} and shell-model effective interaction \cite{Gao99,Hey94}. 
In general, the mean field channel of the np interaction is dominated by its strong quadrupole-quadrupole correlation and strongly attractive  monopole interaction, which are expected to play a key role in
the development of collective correlation~\cite{Fed77,Cas06} and in the evolution of
the shell structure~\cite{Otsuka01,Sor08,Xu13}.

One possible way to extract the residual \emph{np} interaction in odd-odd nuclei (denoted as $\delta_{np}$) is to take the difference between the pairings gaps of even-odd (even nuclei minus one) and even-even nuclei as \cite{Sar15}
\begin{eqnarray}\label{vpn-sc2}
\delta^{(1)}_{np} &=&\Delta^{(3)}_{n} (Z,N) -\Delta_n^{(3)} (Z-1,N) \nonumber \\
&=& \frac{1}{2}\left[
B(Z,N)+B(Z,N-2) \right. \nonumber \\
&&- B(Z-1,N)-B(Z-1,N-2)  \nonumber \\
&&- \left. 2B(Z,N-1)+2B(Z-1,N-1) \right]\nonumber \\
&=&\delta V_{np}(OO)-[2\delta V_{np}(EO)+\delta V_{np} (OE)-2\delta V_{np}(EE)]\nonumber\\
&=&\frac{1}{2}[V_{1n-1p}(Z,N)-V_{1n-1p}(Z,N-1)],
\end{eqnarray}
where $\Delta_n^{(3)}$ denotes the three-point odd-even staggering formula for the empirical neutron pairing gap\footnote{There was a typo in Fig. 2 in Ref. \cite{Sar15} where the $\delta_{pn}$ values plotted correspond to the difference between nuclei with $Z$ and $Z+1$ protons.}. The persistence of positive $\delta_{np}$ values would suggest that $\Delta_n^{(3)}$ for odd-$A$ isotopes is reduced relative to those of the even-even nuclei due to the residual np interaction.
The reduction equals to the difference between $\delta V_{np}(OO)$ and an weighted average of those for odd-$A$ and even-even nuclei. It also equals to half of the difference between $V_{1n-1p}(EE)$ and $V_{1n-1p}(EO)$\footnote{This is pointed out to us by Y.Y. Cheng.}. A very similar result can be obtained by taking the difference between the pairing gaps of the even-even and corresponding even-$Z$-odd-$N$ nucleus.
There can be other ways to extract the residual np interaction \cite{Ber07, Jen84}. One may also simply take the difference between the odd-odd and even-even nuclei as
 \begin{eqnarray}\label{vpn-sc3}
\delta^{(2)}_{np}= \delta V_{np}(OO)-\delta V_{np}(EE),
\end{eqnarray}
which looks simpler than Eq. (\ref{vpn-sc2}) but actually is a seven-point formula involving one more nucleus than Eq. (\ref{vpn-sc2}).
In the spirit of Ref. \cite{Hov13}, one can remove the smooth contribution as defined by the curve for $\delta V_{np}(EE)$ in Fig. \ref{ee-oo} as
 \begin{eqnarray}\label{vpn-sc4}
\delta^{(3)}_{np}=\delta V_{np}(OO)-V_{smooth}.
\end{eqnarray}
It hopes that the $\delta^{(3)}_{np}$ values thus extracted will be less influenced by the local fluctuations presented in the average np interaction in the even-even nucleus in relation to the shell and other effects. However, extra parameters have to be introduced to describe the smooth trend.

\begin{figure}
\includegraphics[width=0.45\textwidth]{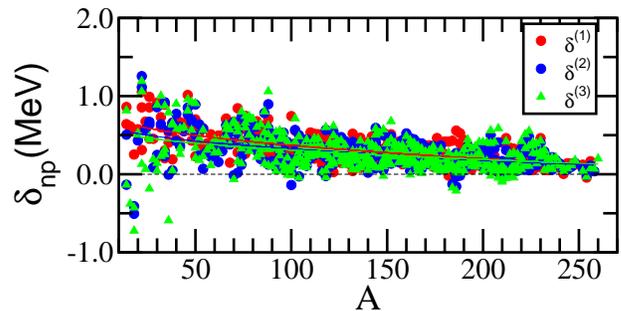}
\caption{\label{csexp} (Color online) The residual np interactions defined in Eqs. (\ref{vpn-sc2}-\ref{vpn-sc4}) for all known nuclei with $N\neq Z$ as extracted from experimental data.}
\end{figure}

In Fig. \ref{csexp} we plotted the residual np interaction extracted from experimental binding energies \cite{AM12} using above three formulas. It is seen that, as expected, the extracted $\delta_{np}$ values are positive for almost all known nuclei. This is particularly true for $\delta_{np}^{(1)}$. The values of all above three formulas show a weak dependence on the mass number $A$. But relative large fluctuations are seen in $\delta_{np}^{(2)}$ and $\delta_{np}^{(3)}$. The mean $\delta_{np}$ values are around 300 keV in all three cases. The values for available superheavy nuclei are mostly below 200 keV.

From Eq. (\ref{vpn-sc2}) one realizes that the positive contribution from the residual np interaction to the total binding energy is the origin of the systematic deviations seen in Figs. \ref{ee-oo} and \ref{eo-oe} as well as the odd-even staggering in $V_{1n-1p}$ that was studied in Ref. \cite{zhao15}. $V_{1n-1p}$ values for even-$A$ nuclei in Fig. \ref{ee-oo} are larger than the mean value of $\delta V_{np}(EE)$ by roughly an amount $\delta_{np}$. Whereas $V_{1n-1p}(EO/OE)$ in Fig. \ref{eo-oe} is smaller than the average by an amount $\delta_{np}$. 

\begin{figure}
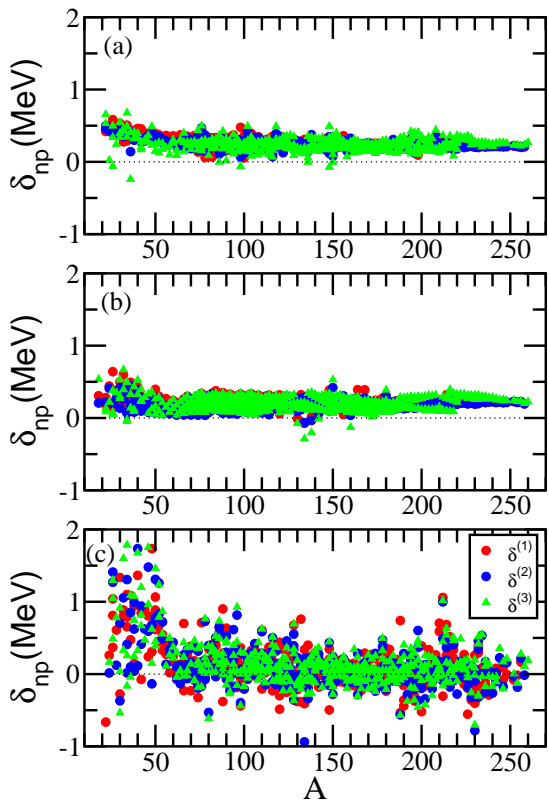

\includegraphics[width=0.4\textwidth]{npeecs3-wang-exp.eps}
\includegraphics[width=0.4\textwidth]{npeecs3-DZ33-exp.eps}
\includegraphics[width=0.4\textwidth]{npeecs3-hfb24-exp.eps}

\caption{\label{csameeya} (Color online) Same as Fig. \ref{csexp} but for those extracted from calculated binding energies from the mac-mic mass formula \cite{wang14} (a), the shell-model mass formulas \cite{Duf95} with parameters as taken from Ref. \cite{Qi15} (b) and the HFB mass formula \cite{hfb24} (c). Only results for experimentally known nuclei are plotted.}
\end{figure}

\subsection{$\delta_{np}$  from recent mass formulas}
$\delta_{np}$ may also serve as a good test to theoretical mass models, in particular to their predictions of the relative evolution of mean field effect in neighboring nuclei, which are supposed to be largely canceled out in $\delta_{np}$. In the upper and middle panels of Fig. \ref{csameeya} we calculated $\delta_{np}$ values from a recent macroscopic-microscopic (mac-mic) mass formula \cite{wang14}  and the 33-term shell-model mass formula \cite{Duf95}.
 As indicated in Fig. \ref{csameeya}, an additional isospin-dependent np interaction term was added to the binding energy of the odd-odd nuclei in the shell-model mass formula \cite{Duf95}. But our calculations show that the improvement in the overall root-mean square error by the addition of this term is marginal \cite{Qi15}. The reason is that large irregular fluctuations around zero occur in intermediate mass and heavy nuclei around pre-defined shell closures. For example, one has $\delta^{(1)}_{np}$ as low as -0.38 MeV for the $N=81$ nucleus $^{132}$Sb in comparison to the experimental data of $\delta^{(1)}_{np}=0.11$ MeV.
This is related to the fact that the binding energies of $^{132}$Sb and the neighboring nucleus $^{133}$Te are underestimated by around 900 and 400 keV, respectively, whereas the binding energy of the semi-magic nucleus $^{133}$Sb is overestimated. Such inconsistent or non-systematic deviations from experimental data lead to large unphysical fluctuations in local np formula $\delta_{np}$. Actually similar fluctuations are also seen in the pairing gaps extracted from the  odd-even staggering in binding energies calculated from the shell-model mass formulas \cite{Sar15a}. 

In the lowest panel of Fig. \ref{csameeya} we plotted the $\delta_{np}$ values calculated and a recent HFB mass formula \cite{hfb24}.
As can be seen from Fig. \ref{csameeya}c, the $\delta_{np}$ values extracted from the HFB mass formula show fluctuations that are significantly larger than those from the mac-mic and shell-model mass formulas. In intermediate and heavy nuclei, the deviation can be as large as $\pm1$ MeV. One has $\delta_{np}>4$MeV in certain light nuclei. For example, for the nucleus $^{18}$F we have $\delta^{(1)}_{np}=5.2$ MeV in comparison to the experimental result of 1.97 MeV. A detailed comparison between the experimental and calculated binding energies of the involved nuclei  shows that the large deviation is related to the underestimation of the binding energy of $^{19}$Ne by 2.3 MeV. As a result, the empirical pairing gap for $^{20}$Ne, which enters Eq. (9), is significantly overestimated.

\begin{figure}
\includegraphics[width=0.45\textwidth]{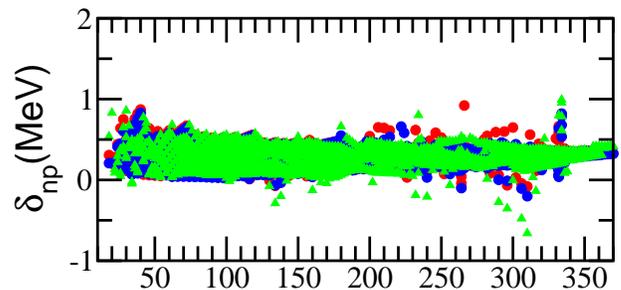}
\caption{\label{all} (Color online) Same as Fig. \ref{csexp} but for all predicted bound nuclei from the full shell-model mass formula.}
\end{figure}

The fluctuations become even larger in dripline nuclei, as seen in Fig. \ref{all} as well as in the $\delta_{np}$ values \cite{supp}. The fluctuations in Fig. \ref{all} occur at mass regions around $A=150$ and 220 (up to $\pm0.5$ MeV) and in unknown heavy and superheavy nuclei. 
Again, the fluctuation is related to the different predictions on the masses of neighboring even and odd mass nuclei around shell closures.
The fluctuations for unknown nuclei as extracted from the HFB mass formula are mostly around $\pm2$ MeV. They are as large as $\pm4$ MeV in a few cases around $A=300$. 
As for those extracted from the mac-mic model, large fluctuations (upto $\pm2$ MeV) also appear in the superheavy region where the masses are not known.

Those large fluctuations seen in Fig. \ref{csameeya}, especially the appearance of large negative $\delta_{np}$ values, are not expected from the systematic behavior as plotted in Fig. \ref{csexp}.
On the other hand, they indicate that the relative evolution of the binding energies of neighboring even and odd nuclei may have not been properly reproduced in a consistent way, even though it may not show itself through the average deviations of the masses between theory and experiment.
The correction to those fluctuations can provide a good constraint on the improvement of the corresponding mass formulas and the underlying shell structure. 

\section{Summary}
To summarize, we compared two families of filters for the extraction of the average np interaction from nuclear binding energies, which have been studied intensively recently. All these formulas involve four neighboring nuclei. One avoids breaking the proton and/or neutron pair in the construction the first family of np formulas [Eqs. (1-4)]. As for odd-odd nuclei, the second family, $V_{1n-1p}$ in Eq. (5), coincides with the first type Eq. (4). $V_{1n-1p}$ values for even-even and odd-odd systems are similar to each other quantitatively and are systematically \textit {larger} than $\delta V_{np}(EE)$. On the other hand, $V_{1n-1p}$ values for odd-$A$ nuclei are systematically \textit{ smaller} than those for neighboring even-$A$ systems as well as than $\delta V_{np}(EE)$.
These systematic deviations are related to the fact that there is an additional np interaction present in the odd-odd nuclei in comparison to the neighboring even-even and odd-$A$ systems [Eq. (\ref{vpn-sc2})]. 

We explored three possible ways to extract the additional np interaction from the binding energy difference of neighboring nuclei [Eqs. (\ref{vpn-sc2}-\ref{vpn-sc4})]. The  extracted $\delta_{np}$ values are positive in nearly all cases, as expected, and show a weak dependence on mass number $A$. We then compared the empirical $\delta_{np}$ values with those extracted from recent nuclear mass models. Large fluctuations around zero ($\pm1$ MeV for known nuclei and $\pm2$ MeV for unknown nuclei) are seen in the HFB mass model.
 Fluctuations  are also seen in the shell-model mass formulas at mass regions around $A=150$ and 220 (up to $\pm0.5$ MeV) and in unknown superheavy nuclei (up to $\pm1$ MeV). 
The irregular fluctuations are related to the different behaviors in the predicted binding energies of the neighboring even and odd mass nuclei involved. The residual np interaction $\delta_{np}$ may serve as an excellent criterion for the refinement of those mass models and for the constraint on the underlying shell structure.

\section*{Acknowledgement}
This work was supported by the Swedish Research Council (VR) under grant Nos. 621-2012-3805, 621-2013-4323 and the Jiangsu overseas research and training program for university prominent young and middle-aged teachers. CQ thanks the Swedish National Infrastructure for Computing (SNIC) at NSC in Link\"oping and PDC at KTH, Stockholm for computational support and the nuclear physics group at Shanghai Jiaotong University for their hospitality.


\begin{thebibliography}{50}
\bibitem{Hov13}D. Hove, A. S. Jensen, and K. Riisager, Phys. Rev. C {\bf88}, 064329 (2013).
\bibitem{Satu98}W. Satu{\l}a, J. Dobaczewski, and W. Nazarewicz, Phys. Rev. Lett. {\bf81}, 3599 (1998).
\bibitem{Ber09}G. F. Bertsch, C. A. Bertulani, W. Nazarewicz, N. Schunck, and M. V. Stoitsov,
Phys. Rev. C {\bf79}, 034306 (2009).
\bibitem{Sar15}S. A. Changizi and C. Qi, Phys. Rev. C  {\bf91}, 024305 (2015).
\bibitem{Sar15a}S.A. Changizi, C. Qi, and R. Wyss,  Nucl. Phys. A {\bf940}, 210 (2015).

\bibitem{Zhang89} J.-Y. Zhang, R. F. Casten and D. S. Brenner,
Phys. Lett. B {\bf227}, 1 (1989).
\bibitem{Cas05}R. B. Cakirli, D. S. Brenner, R.F. Casten and E.A. Millman,
Phys. Rev. Lett. {\bf94}, 092501 (2005).
\bibitem{Cas06}R. B. Cakirli and R.F. Casten, Phys. Rev. Lett. \textbf{96}, 132501 (2006).
\bibitem{Cas06a}Y. Oktem, R. B. Cakirli, R. F. Casten, R. J. Casperson, and D. S. Brenner,
Phys. Rev. C {\bf74}, 027304 (2006).
\bibitem{Cas06b}D. S. Brenner, R. B. Cakirli, and R. F. Casten, Phys. Rev. C {\bf73}, 034315 (2006).
\bibitem{Isa95}P. Van Isacker, D.D. Warner, and D.S. Brenner, Phys. Rev. Lett. {\bf74}, 4607 (1995).
\bibitem{Satu97} W.~Satu{\l}a, D.J. Dean, J. Gary, S. Mizutori, W. Nazarewicz, Phys.~Lett.~{\bf B407}, 103 (1997).
\bibitem{Chen09}L. Chen {\it et al.}, Phys. Rev. Lett. {\bf102}, 122503 (2009).
\bibitem{Bon13}D. Bonatsos, S. Karampagia, R. B. Cakirli, R. F. Casten, K. Blaum, and L. A. Susam, Phys. Rev. C {\bf88}, 054309 (2013).
\bibitem{Cas09}R. B. Cakirli, R. F. Casten, R. Winkler, K. Blaum, and M. Kowalska, Phys. Rev. Lett. {\bf102}, 082501 (2009).
\bibitem{Chas07}R.R. Chasman, Phys. Rev. Lett. {\bf99}, 082501 (2007).
\bibitem{Isa07}A.E.L. Dieperink, and P. Van Isacker, Eur. Phys. J. A {\bf32}, 11 (2007).
\bibitem{Sto07}M. Stoitsov, R. B. Cakirli, R. F. Casten, W. Nazarewicz, and W. Satu{\l}a,
Phys. Rev. Lett. {\bf98}, 132502 (2007).
\bibitem{Qi12} C. Qi, Phys. Lett. B {\bf717}, 436 (2012).
\bibitem{Bas71}M. K. Basu and D. Banerjee, Phys. Rev. C {\bf3}, 992 (1971).
\bibitem{zhao15}Y. Y. Cheng, Y. M. Zhao, and A. Arima, Phys. Rev. C {\bf91}, 024314 (2015).
\bibitem{zhao11}G. J. Fu, Y. Lei, H. Jiang, Y. M. Zhao, B. Sun, and A. Arima, Phys. Rev. C {\bf84}, 034311 (2011).
\bibitem{zhao12}A. Arima and Y.M. Zhao,
J. Phys.: Conf. Ser. 403, 012045 (2012).
\bibitem{zhao13}G. Fu, J. Shen, Y. Zhao, and A. Arima,
Phys. Rev. C 87, 044309 (2013).
\bibitem{Zhao14}Y. Lu, Y. M. Zhao, and A. Arima, Phys. Rev. C 89, 017301 (2014).

\bibitem{AM12} M. Wang, G. Audi, A. H. Wapstra, F. G. Kondev et al., Chin.
Phys. C {\bf36}, 1603 (2012).
\bibitem{Wu15} Z.Y. Wu, C. Qi, R. Wyss and H.L. Liu,  Phys. Rev. C 92, 024306 (2015).
\bibitem{Macc00}A.O. Macchiavelli {\textit et al.}, Phys. Rev. C {\bf61}, 041303(R) (2000).
\bibitem{Qi11}C. Qi, J. Blomqvist, T. B\"ack, B. Cederwall, A. Johnson, R. J. Liotta, and R. Wyss,
Phys. Rev. C {\bf84}, 021301(2011); B. Cederwall et al., Nature 469, 68 (2011); C. Qi and R. Wyss, Physica Scripta 91, 013009 (2016).
\bibitem{Good01}A.L. Goodman, Phys. Rev. C {\bf63}, 044325 (2001).
\bibitem{Satu01}W. Satu{\l}a and R. Wyss, Phys. Rev. Lett. {\bf86}, 4488 (2001);
Phys. Rev. Lett. {\bf87}, 052504 (2001).
\bibitem{Ber07}W. A. Friedman and G. F. Bertsch, Phys. Rev. C  {\bf76}, 057301 (2007).
\bibitem{Gao99}Z. C. Gao and Y. S. Chen, Phys. Rev. C {\bf59}, 735 (1999).
\bibitem{Hey94}K. Heyde, C. De Coster and J. Schietse, Phys. Rev. C {\bf49}, 2499 (1994).

\bibitem{Fed77}P. Federman and S. Pittel, Phys. Lett. B {\bf 69},  385 (1977).
\bibitem{Otsuka01}T. Otsuka, R. Fujimoto, Y. Utsuno, B. A. Brown, M. Honma, and T. Mizusaki,
Phys. Rev. Lett. {\bf87}, 082502 (2001).
\bibitem{Sor08}O. Sorlin, M.G. Porquet, Prog. Part. Nucl. Phys. {\bf61}, 602 (2008).
\bibitem{Xu13}Z. Xu and C. Qi, Phys. Lett. B, 724, 247 (2013).

\bibitem{Jen84}A. S. Jensen, P. G. Hansen, and B. Jonson, Nucl. Phys. A 431, 393 (1984).
\bibitem{wang14}N. Wang, M. Liu, X. Wu, and J. Meng, Phys. Lett. B {\bf734}, 215 (2014).
\bibitem{Duf95}J. Duflo and A.P. Zuker, Phys. Rev. C {\bf52}, R23 (1995).
\bibitem{Qi15}C. Qi, J. Phys. G: Nucl. Part. Phys. 42, 045104 (2015).
\bibitem{hfb24}S. Goriely, N. Chamel, and J. M. Pearson, Phys. Rev. C {\bf88}, 024308 (2013).
\bibitem{supp}See Supplemental Material at [URL will be inserted by publisher]
for $\delta_{np}$ values extracted from experimental and theoretical binding energies.
\end{thebibliography}
\end{document}